\documentclass[aps,prl,reprint,superscriptaddress,floatfix]{revtex4-2}
\usepackage{amsmath,amssymb,wasysym}
\usepackage{graphicx}
\usepackage{dcolumn}
\usepackage{bm}
\usepackage{braket}
\usepackage{verbatim}
\usepackage{float}
\usepackage{bbold}
\usepackage{color}
\usepackage{xcolor}
\usepackage{relsize}
\usepackage{amsthm}
\usepackage{enumerate}
\usepackage{soul,xcolor}
\usepackage[T1]{fontenc}
\usepackage{adjustbox}
\usepackage[colorlinks=true ,urlcolor=blue,urlbordercolor={0 1 1}]{hyperref}

\renewcommand{\[}{\left[}

\newcommand{\be}{\begin{equation}}
\newcommand{\ba}{\begin{align}}
\newcommand{\ee}{\end{equation}}
\newcommand{\bea}{\begin{eqnarray}}
\newcommand{\eea}{\end{eqnarray}}
\newcommand{\beq}{\begin{equation}}
\newcommand{\eeq}{\end{equation}}
\newcommand{\beqn}{\begin{eqnarray}}
\newcommand{\eeqn}{\end{eqnarray}}

\usepackage{bm}
\usepackage{array}
\newcolumntype{L}[1]{>{\raggedright\arraybackslash}p{#1}}
\newcolumntype{C}[1]{>{\centering\arraybackslash}p{#1}}
\newcolumntype{R}[1]{>{\raggedleft\arraybackslash}p{#1}}
\usepackage{multirow}
\allowdisplaybreaks

\begin{document}
\title{Incommensurate pair-density-wave correlations in two-leg ladder $t$--$J$--$J_\perp$ model}

\author{Hanbit Oh}
\thanks{Corresponding author: \href{mailto:hoh22@jh.edu}{hoh22@jh.edu}}
\affiliation{William H. Miller III Department of Physics and Astronomy, Johns Hopkins University, Baltimore, Maryland, 21218, USA}
\author{Julian May-Mann}
\affiliation{Department of Physics, Stanford University, Stanford, CA 94305, USA}
\author{Ya-Hui Zhang}
\thanks{Corresponding author: \href{mailto:yzhan566@jhu.edu}{yzhan566@jhu.edu}}
\affiliation{William H. Miller III Department of Physics and Astronomy, Johns Hopkins University, Baltimore, Maryland, 21218, USA}

\date{\today}

\begin{abstract}
We report the discovery of a generalized Luther-Emery liquid phase characterized by incommensurate pair-density-wave (iC-PDW) correlations in the two-leg $t$-$J$-$J_\perp$ ladder model. By tuning the potential difference between the legs, we explore the regime of intermediate layer polarization $P$. Combining density-matrix renormalization group (DMRG) simulations with bosonization analysis, we identify a spin-gapped phase at finite $P$, where the interlayer and intralayer pair correlations both oscillate, but with distinct periodicities. The interlayer correlations exhibit FFLO-like oscillations, driven by pairing between layers with mismatched Fermi momenta, with a period determined by their momentum difference. In contrast, the intralayer pair correlations arise from the coupling between charges on one layer and spin fluctuations on the opposite layer, with a momentum equal to twice the Fermi momentum of the opposite layer. The iC-PDW state is robust across a wide range of doping and polarization, although finite interlayer hopping eventually destabilizes it toward a state with charge-$4e$ correlations. We conclude by discussing the experimental realization of this model in optical lattice platforms and its relevance to the bilayer nickelate La$_3$Ni$_2$O$_7$. 
\end{abstract}

\maketitle
{\it Introduction.---} 
Pair-density-wave (PDW) superconductivity, characterized by Cooper pairing at finite center-of-mass momentum, has been one of the central themes in the study of strongly correlated systems~\cite{agterberg2020physics}. 
A notable example of finite-momentum pairing is the Fulde--Ferrell--Larkin--Ovchinnikov (FFLO) phase, where pairing is driven by an external Zeeman field~\cite{fulde1964superconductivity,larkin1964nonuniform}. 
On the other hand, a time-reversal-invariant PDW needs an intrinsically strong-coupling pairing mechanism to compete with the uniform superconductor. One well-established realization is in the one-dimensional Kondo--Heisenberg model, which exhibits a commensurate PDW at momentum $q=\pi$ together with a finite spin gap~\cite{Sikkema1997Spin,Berg2010Pair}.
There is also a wide range of theoretical and numerical studies which have investigated PDW phases in various microscopic models~\cite{Cho2014Topological,May-Mann2020Topology,Zhang2022Pair,Nikolaenko2024Numerical,Huang2024Enhanced,Huang2022Pair-density-wave,Jiang2024Pair-Density-Wave,Chen2024Pair,Liu2024Pair,Chen2023Singlet,Jiang2023Pair,Peng2021Gapless,Venderley2019Evidence,Xu2019Pair,Venderley2019Density,Soto-Garrido2015Quasi-one-dimensional, yang2024phase}. 
Meanwhile, incommensurate PDW phases are less explored. To our knowledge, incommensurate PDW correlations have not yet been unambiguously demonstrated even in numerical studies.

In this work, we propose a two-leg ladder $t$--$J$--$J_\perp$ model as an ideal platform to search for incommensurate pair-density-wave (iC-PDW) correlations. The model consists of two $t$--$J$ chains coupled solely through an interlayer spin-exchange interaction $J_\perp$, in the absence of interlayer charge hopping $t_\perp$ [Fig.~\ref{fig:1}(a)]. At zero layer-polarization, it hosts a uniform interlayer superconducting state with dominant singlet pairing correlations, corresponding to a Luther--Emery liquid in one dimension~\cite{Luther1974Backward}. Beyond one dimension, the same model has been widely studied in the context of the bilayer nickelate La$_3$Ni$_2$O$_7$~\cite{sun2023signatures}, where strong Hund’s coupling between the $d_{x^2-y^2}$ and $d_{z^2}$ orbitals generates an effective interlayer spin exchange.
In this setting, previous theoretical studies have revealed that the ground state is an interlayer $s$-wave superconductor~\cite{Oh2023Type-II,Qu2024Bilayer,lu2023superconductivitydopingsymmetricmass,Lu2024Interlayer,Yang2024Strong,Lange2024Pairing,schlömer2024superconductivity,yang2025strong,wang2025originspinstripesbilayer,oh2025hightemperature,oh2025doping}. Most previous studies are limited to the mirror-symmetric regime of equal layer densities, and the effects of symmetry breaking have been relatively less explored~\cite{fan2025minimal,oh2025pairdensitywave}.

\begin{figure}[b]
    \centering
\includegraphics[width=.8\linewidth]{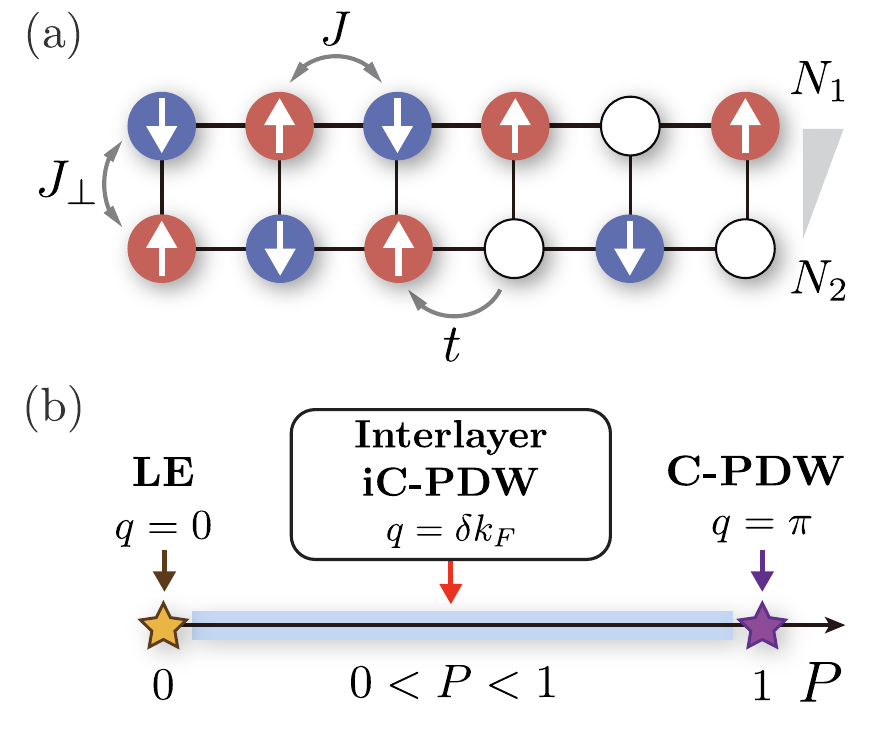}
\vspace{-5pt}
    \caption{(a) Illustration of the two-leg ladder $t$–$J$–$J_\perp$ model. Particle numbers on the two legs are independently conserved, and we set $N_1 \ge N_2$.
(b) Polarization-imbalance–driven phase transitions at $t \sim J \sim J_\perp$.
For $0 < P < 1$, the system hosts an interlayer iC-PDW as the dominant correlation,
with pairing momentum $q = \delta k_F = |k_{F,1} - k_{F,2}|$.
We also find intralayer pairing on layer~2 as a subdominant correlation,
with a distinct ordering wave vector $q = 2k_{F,1}$.
At $P = 0$, the ground state is a Luther–Emery (LE) liquid,
while at $P = 1$ the model yields a commensurate PDW with $q = \pi$.
}
    \label{fig:1}
\end{figure}

Here, we study the $t$-$J$-$J_\perp$ model partially polarized along the rung direction, corresponding to a population imbalance between the two legs. Combining Abelian bosonization~\cite{giamarchi2003quantum,fradkin2013field} and density-matrix renormalization group (DMRG) calculations~\cite{White1992Density,Hauschild2018Efficient}, we report a generalized Luther-Emery liquid phase that hosts two distinct types of iC-PDW correlations: an FFLO-like \textit{interlayer} PDW and a Kondo-Heisenberg-like \textit{intralayer} PDW. The interlayer PDW arises because each ladder leg has a well-defined Fermi momentum due to charge conservation per leg, and a finite polarization causes a mismatch between these Fermi momenta. As in the standard FFLO argument~\cite{fulde1964superconductivity,larkin1964nonuniform}, interlayer pairing can then induce finite-momentum superconducting correlations. 
The intralayer PDW can be understood by considering the limit where one of the legs is half-filled, in which the $t$-$J$-$J_\perp$ model is equivalent to the generalized Kondo model~\cite{Zhang2022Pair}\footnote{The generalized Kondo model slightly departs from the standard Kondo-Heisenberg model~\cite{Sikkema1997Spin} by restricting double occupancy.}. The ground state of the generalized Kondo model hosts $q=\pi$ intralayer commensurate PDW (C-PDW) correlations with a finite spin gap~\cite{Zhang2022Pair}, realizing the same phase found in the standard Kondo-Heisenberg model~\cite{Berg2010Pair}. The incommensurate intralayer correlations of the $t$-$J$-$J_\perp$ model are understood as descendants of these commensurate correlations. We numerically demonstrate that both types of PDW correlations in the $t$-$J$-$J_\perp$ model are robust over a wide range of parameters. However, a small interlayer tunneling $t_\perp$ induces an instability of the PDW phase toward a phase with charge-$4e$ correlations~\cite{berg2009charge, agterberg2008dislocations, wu2024d, huecker2026vestigial}.

Our results establish the $t$-$J$-$J_\perp$ ladder as a unified platform for realizing distinct pairing regimes upon tuning the polarization: a uniform Luther-Emery liquid~\cite{Luther1974Backward} at $P=0$, an interlayer iC-PDW for $0 < P < 1$, and a C-PDW at full polarization $P=1$ [Fig.~\ref{fig:1}(b)]. The model can be simulated in optical lattices~\cite{bohrdt2022strong,hirthe2023magnetically}, where the polarization is easily controllable. While the zero-polarization case has been experimentally explored~\cite{hirthe2023magnetically}, our results provide a strong motivation for future experiments targeting the finite-polarization regime.

\textit{Model.---}
We consider the two-leg ladder $t$--$J$--$J_\perp$ model Hamiltonian, illustrated in Fig.~\ref{fig:1}~(a),
\begin{align}
H &= 
-t\sum_{\ell,\sigma}\sum_{\langle i,j\rangle}
\mathcal{P}\left(
c^{\dagger}_{i,\ell,\sigma} c_{j,\ell,\sigma}
\right)\mathcal{P} + \mathrm{H.c.} \notag \\
&\quad 
+ J \sum_{\ell}\sum_{\langle i,j\rangle} 
\vec{S}_{i,\ell}\cdot \vec{S}_{j,\ell}
+ J_\perp \sum_{i} 
\vec{S}_{i,1}\cdot \vec{S}_{i,2},
\label{eq:t-J-Jp}
\end{align}
where $i$ labels the rung index and $\langle i,j\rangle$ runs over nearest neighbors along each leg.
The leg index is $\ell=1,2$ and the spin index is $\sigma=\uparrow,\downarrow$.  
Here $c^{\dagger}_{i\ell\sigma}$ creates an electron with quantum numbers $(i,\ell,\sigma)$,  
$n_{i\ell\sigma}=c^{\dagger}_{i\ell\sigma}c_{i\ell\sigma}$, and 
$\vec{S}_{i\ell}=\frac{1}{2}c^{\dagger}_{i\ell\sigma}\vec{\sigma}_{\sigma\sigma'}c_{i\ell\sigma'}$.
The projection operator $\mathcal{P}$ enforces no double occupancy on each site.

The Hamiltonian conserves $(N_1,N_2)$, the total particle numbers on each leg, and the total spin $S$, reflecting a $(U(1)\times U(1)\times SU(2))/Z_2$ symmetry.  
We consider a situation starting from a Mott insulator at half filling in both layers and then dope holes into each layer.
For convenience, we define the hole concentration as $h_\ell = 1 - n_\ell$, with $n_\ell = N_\ell/L$.
The average filling is given by $n = (n_1 + n_2)/2 = 1 - h$.
We introduce a normalized polarization to quantify the imbalance of hole densities,
\begin{equation}
    P = \frac{n_1 - n_2}{2(1 - n)}
      = \frac{h_2 - h_1}{2h}.
\end{equation}
Without loss of generality, we assume that more holes enter to layer~2, i.e., $h_1 \le h_2$
(equivalently $n_2 \le n_1$).
With this parametrization, $P=1$ corresponds to the fully polarized limit in which all doped holes enter to layer~2,
while layer~1 remains at half filling.
Throughout this work, we mainly consider $n=7/8$ and set $t=1$, $J_\perp=2$ and vary $J$ and $P$.
Our goal is to determine how the ground state evolves with polarization $P$, especially superconducting correlations. As we demonstrate below, the partially polarized regime hosts an iC-PDW state with a finite spin gap with central charge $c=2$

Note that the model includes an interlayer exchange coupling $J_\perp$ but excludes interlayer hopping terms $t_\perp$. In the context of bilayer nickelates, this is reasonable, as the effective $J_\perp$ of the $d_{x^2-y^2}$ orbitals is mediated by a Hund's coupling between the itinerant $d_{x^2-y^2}$ orbitals and the localized $d_{z^2}$ orbitals~\cite{Oh2023Type-II}. It may also be possible to realize Eq.~\ref{eq:t-J-Jp} in cold-atom quantum simulators via an out-of-equilibrium implementation known as the mixed-dimensional (mixed-$D$) $t$-$J$ model~\cite{bohrdt2022strong,hirthe2023magnetically}. While this model can be generalized to any lattice, here we focus on the one-dimensional ladder geometry.

\textit{Numerical Simulation.---}
We perform density-matrix renormalization group (DMRG) simulations of Eq.~\eqref{eq:t-J-Jp} using the TeNPy package~\cite{Hauschild2018Efficient}.
Open boundary conditions are imposed along the chain direction, and system sizes
$L=64$--$128$ are considered.
The bond dimension is varied in the range $\chi=2000$--$4000$, depending on parameters,
with typical truncation errors below $10^{-6}$.
For computational efficiency, we exploit three conserved $U(1)$ charges $(N_1, N_2, S_z)$.

\begin{figure}[tb]
    \centering
\includegraphics[width=\linewidth]{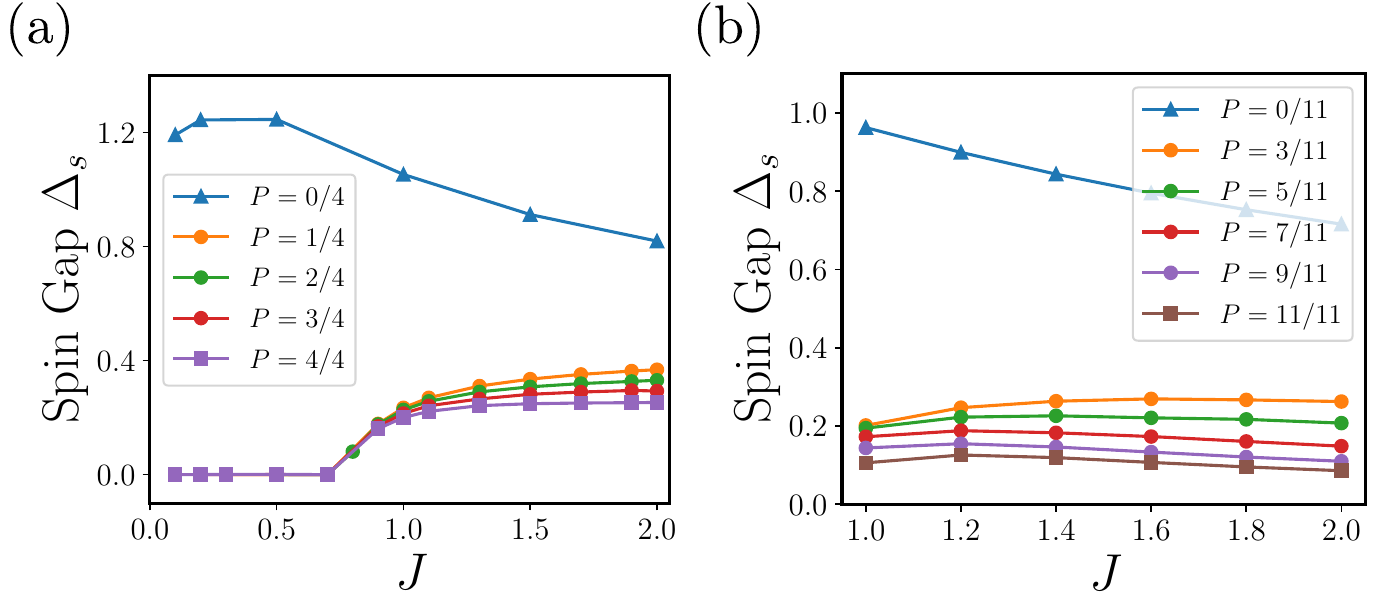}
    \caption{
    Finite DMRG simulation of the $t$--$J$--$J_\perp$ model. We use parameters $t=1$ and $J_\perp=2$ with (a) $n=7/8$ and (b) $n=38/47$.
(a,b) Spin gap as a function of $J$ for various polarizations $P$. 
Simulations are performed on a system of length (a) $L=64$ and (b) $L=94$ with bond dimension $\chi=3000$. 
(a) While the spin gap closes for $J<J_c=0.78$, our main interest lies in the spin-gapped regime $J>J_c$, where the spin-gapped iC-PDW is realized.
}
    \label{fig:2}
\end{figure}

The main finite-DMRG results are summarized in Fig.~\ref{fig:2} and Fig.~\ref{fig:3}. 
Fig.~\ref{fig:2}(a) shows the spin gap, $\Delta_s = E(S_z=1) - E(S_z=0)$.
For sufficiently large $J >J_c=0.78$, a finite spin gap persists across the entire polarization range $0 \le P \le 1$. Although Fig.~\ref{fig:2}(a) only shows the cases $n=7/8$ and $n=38/47$, the spin gap is robust over different fillings and generally increases with increasing electron density, as shown in the Supplemental Material (SM). 
By contrast, in the small-$J$ regime, $J < J_c = 0.78$ in Fig.~\ref{fig:2}(a), we find that the spin gap disappears for any nonzero polarization $P \neq 0$. 
The transition is first order. Hereafter, we mainly focus on the spin-gapped phase at large $J$.

There are two types of pairing channels to consider here:
\begin{itemize}
    \item \textit{Interlayer}: $\Delta_\perp(i) = c_{i,1,\uparrow} c_{i,2,\downarrow} - c_{i,1,\downarrow} c_{i,2,\uparrow}$,
    \item \textit{Intralayer}: $\Delta_{\parallel,\ell}(i) = c_{i,\ell,\uparrow} c_{i+1,\ell,\downarrow} - c_{i,\ell,\downarrow} c_{i+1,\ell,\uparrow}$,
\end{itemize}
with leg index $\ell=1,2$.
In Figs.~\ref{fig:3}(a) and (b), we show the real-space pair-correlation functions at (a) $P=2/4$, $n=7/8$ and (b) $P=3/11$, $n=38/47$. Both interlayer and intralayer pairing correlations exhibit algebraic decay. By extracting the corresponding power-law exponents, we find the clear hierarchy: $\alpha_\perp < \alpha_{\parallel,2} < \alpha_{\parallel,1}$ for the interlayer, leg-2 intralayer, and leg-1 intralayer pairs, respectively. Consequently, the interlayer PDW is identified as the dominant quasi-long-range order in this regime. Within the intralayer channels, the pairing correlations on the less populated leg $\Delta_{\parallel,2}$ decay significantly more slowly than those on the more populated leg $\Delta_{\parallel,1}$.

Figs.~\ref{fig:3}(c-f) show the Fourier transforms of the pair correlation functions. 
For partial polarization $0<P<1$, the interlayer pairing channel exhibits a sharp peak at 
$q=\delta k_F = |k_{F,1}-k_{F,2}|$, consistent with finite-momentum pairing driven by Fermi-momentum mismatch.  The behavior of intralayer pairing, however, is strongly layer dependent. For the less populated layer~2, the intralayer pairing shows a clear peak at momentum $2k_{F,1}$ [Figs.~\ref{fig:3}(e,f)], reflecting $2k_{F}$ signal from the opposite layer. 
By contrast, the intralayer pairing correlations on layer~1 do not exhibit any sharp finite-momentum peak [Fig.~\ref{fig:s3}].  
These finding can be understood by viewing the intralayer pairing on layer~$\ell$ as a composite order parameter, formed by the uniform triplet superconducting pairing on layer~$\ell$ and the $2k_F$ spin-density-wave fluctuations on the opposite layer~$\overline{\ell}$.  Within this scenario, we can explain why the induced intralayer pairing correlations on layer~1 are weaker: since layer~2 is further away from half filling than layer~1,
its spin fluctuations are weaker, which in turn suppresses the induced intralayer PDW correlations on layer~1.

For comparison, we also check the limiting cases $P=0$ and $P=1$. At $P=0$, only interlayer zero-momentum pairing appears, consistent with the Luther–Emery phase.  
At $P=1$, the leg 1 becomes half-filled, and the remaining intralayer pairing shows a peak at $q=\pi$, in agreement with the analysis of the generalized Kondo model in Ref.~\cite{Zhang2022Pair}.

\begin{figure}[t]
    \centering
\includegraphics[width=\linewidth]{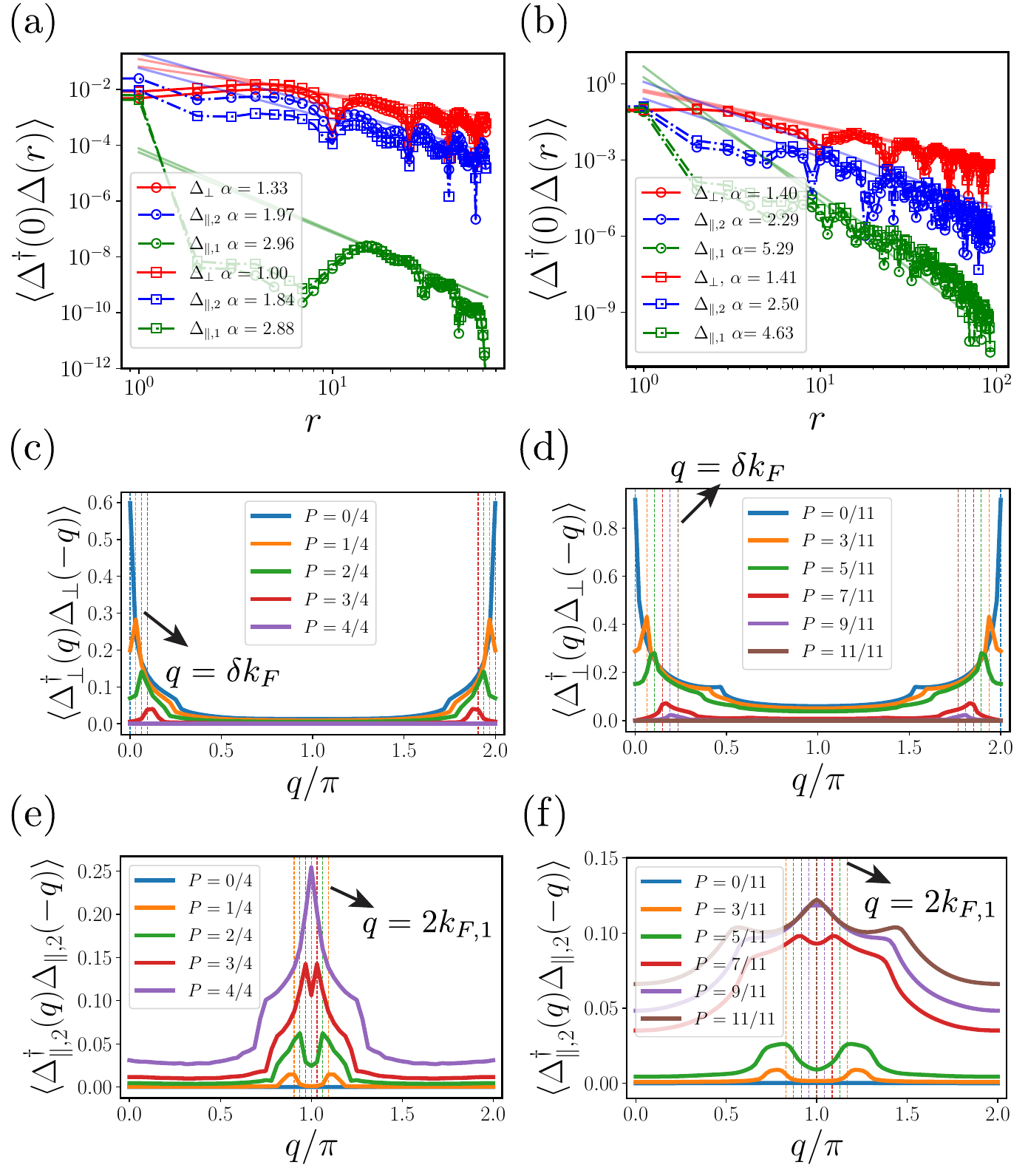}
    \caption{
    Pair correlations of interlayer pair and intralayer pair. 
    We use parameters $t=1$, $J_\perp=2$ with (a,c,e) $n=7/8$ and (b,d,e) $n=38/47$.
    (a) Real space pair correlation shows power-law decaying at
    $P=2/4$ with $J=1$~(circle) and $J=1.4$~(square), (b) $P=3/11$ with $J=1$~(circle) and $J=1.5$~(square). 
$\alpha$ is the power-law exponent and we find $\alpha_\perp<\alpha_{\parallel,2}<\alpha_{\parallel,1}$ indicates the interlayer pair correlations are dominant long-range correlation.  
(c,d) The interlayer pairing $\Delta_{\perp}$ shows a peak at 
$q=|k_{F,1} - k_{F,2}|=(1-n)P\pi$, denoted as dashed line. (e,f) The intralayer pairing $\Delta_{\parallel,2}$ exhibits a peak at $q=2k_{F,1}=n_1 \pi$, the Fermi momentum of layer 1, denoted as dashed line. 
This peak is sharp at large $P$, where $n_1$ (carrier of spin-fluctuation) is large, but becomes broadened as $P$ decreases.
In contrast, $\Delta_{\parallel,1}$ does not exhibit a sharp peak for any value of $P$ (see Fig.~\ref{fig:s3}).
In evaluating $\langle \Delta^\dagger(0)\Delta(r)\rangle$, we omitted $|r|=0,1$. 
}
    \label{fig:3}
\end{figure}

As a complementary approach, we present infinite DMRG (iDMRG) results in Fig.~\ref{fig:4}. The entanglement entropy scaling shown in Fig.~\ref{fig:4}(a) yields a central charge of $c=2$ for the intermediate-polarization regime, indicating two gapless charge modes for the iC-PDW phase. Furthermore, we utilize the transfer-matrix technique to evaluate the correlation lengths ($\xi$) of various operators within their respective symmetry sectors. As illustrated in Fig.~\ref{fig:4}(b), the correlation lengths for the single-electron and spin sectors remain finite, indicating that these excitations are gapped. In contrast, the correlation lengths for the pair operators  are divergent, indicating a gapless charge 2e excitation. These findings are in excellent agreement with the finite DMRG results of various physical operators (see Fig.~\ref{fig:s5}).

\begin{figure}[tb]
    \centering
\includegraphics[width=\linewidth]{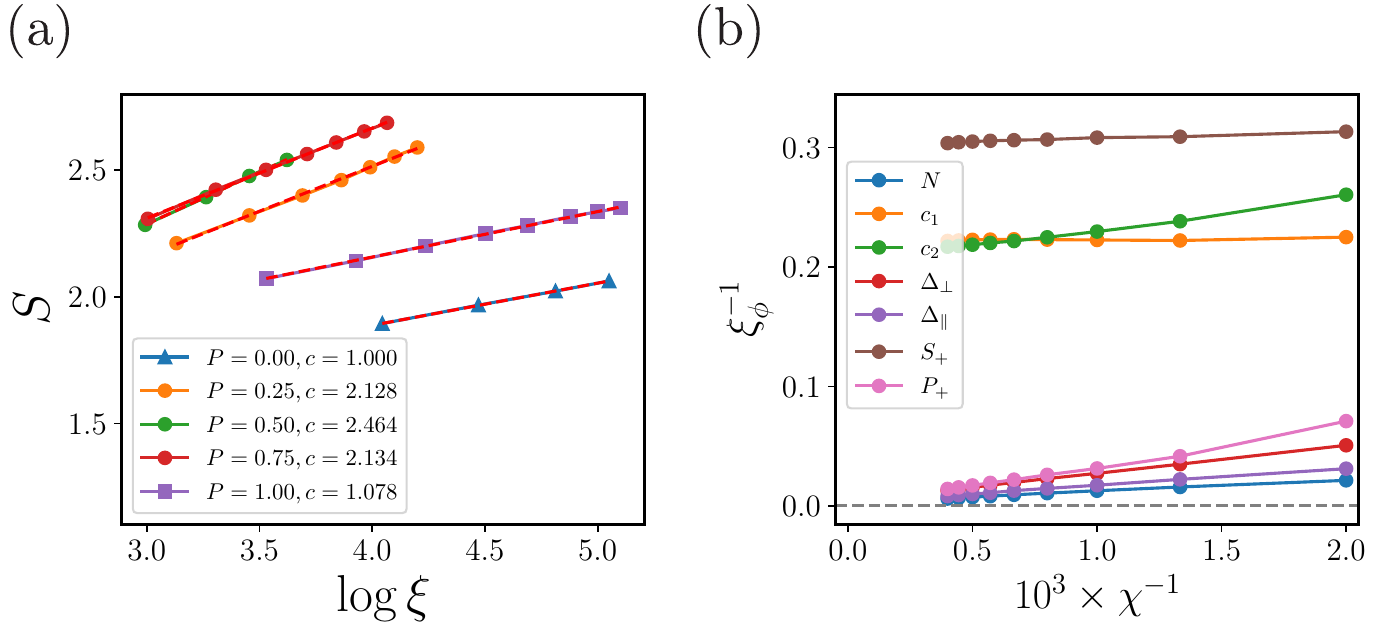}
\caption{
    Infinite DMRG results at filling $n=7/8$ with parameters $t=1$ and $J=J_\perp=2$. 
(a) Entanglement entropy $S$ versus correlation length $\xi$.
The central charge is extracted from the relation $S = \frac{c}{6}\log \xi$. 
In the intermediate-polarization regime $0<P<1$, corresponding to the iC-PDW phase, we find $c=2$, whereas $c=1$ for both the $P=0$ Luther–Emery phase and the $P=1$ commensurate PDW (C-PDW) phase. 
(b) Correlation lengths of different operators, computed using the transfer-matrix technique.
The corresponding charge sectors are specified in terms of $(\delta N_1,\delta N_2,2\delta S_z)$ as 
$(0,0,0)$, $(1,0,1)$, $(0,1,1)$, $(1,1,0)$, $(0,2,0)$, $(0,0,2)$, and $(1,-1,0)$ for 
$N$, $c_t$, $c_b$, $\Delta_\perp$, $\Delta_\parallel$, $S_+$, and $P_+$, respectively. Here, $P_+$ is an operator raising $N_1$ and lowering $N_2$ by 1. 
}
    \label{fig:4}
\end{figure}

\textit{Bosonization Analysis.---}
To analytically clarify the nature of the PDW phase, we employ bosonization~\cite{giamarchi2003quantum,fradkin2013field}, a controlled and powerful framework for one-dimensional correlated systems. Our treatment follows the convention of Ref.~\cite{giamarchi2003quantum}, and we perform an analysis similar to those in Refs.~\cite{Jaefari2012Pair,Zhang2022Pair}.
We begin from the decoupled limit $J_\perp=0$, where each leg hosts gapless charge and spin modes, totaling four gapless modes. The low-energy physics is described by bosonic fields $(\phi_{\ell,c}, \theta_{\ell,c})$ and $(\phi_{\ell,s}, \theta_{\ell,s})$ for charge and spin sectors, respectively. 
For simplicity, we here assume that the Luttiner parameter is the same for each layer, $K_{\ell,c} = K_{c}$.
Once the legs are coupled, it is convenient to transform to the symmetric (even) and antisymmetric (odd) sectors: $\phi_{\pm,\alpha} = (\phi_{1,\alpha} \pm \phi_{2,\alpha})/\sqrt{2}$ and $\theta_{\pm,\alpha} = (\theta_{1,\alpha} \pm \theta_{2,\alpha})/\sqrt{2}$ for $\alpha=c,s$. The Hamiltonian for the coupled ladder is given by
\begin{equation}
\begin{split}
\mathcal{H} 
&= \sum_{\sigma \in \{+,-\}} \frac{v_{\sigma,c}}{2\pi} \left[ K_{\sigma,c}(\partial_x \theta_{\sigma,c})^2 + \frac{1}{K_{\sigma,c}}(\partial_x \phi_{\sigma,c})^2 \right] \\
&\quad + \sum_{\sigma \in \{+,-\}} \frac{v_{\sigma,s}}{2\pi} \left[ (\partial_x \theta_{\sigma,s})^2 + (\partial_x \phi_{\sigma,s})^2 \right] + \mathcal{H}_{\text{int}}.
\end{split}
\end{equation}
Here, $K_{\pm,c}$ represents the Luttinger parameters for the even and odd charge sectors, while $K_{\pm,s}$  denotes those for the spin sectors, typically fixed to $1$ in the presence of $SU(2)$ symmetry. These parameters are determined by $t$, $J$, $J_\perp$, and the filling $n_\ell$. The variables $v_{\pm,c}$ and $v_{\pm,s}$ correspond to the charge and spin velocities for the even and odd sectors, respectively.

$\mathcal{H}_{\text{int}}$ contains the non-trivial interactions in the bosonized theory. Bosonizing the $J_\perp$ term and retaining the most relevant contribution yields 
\begin{equation}
\mathcal{H}_{\text{int}} \simeq - g \cos(\sqrt{2\pi}\phi_{+,s})\cos(\sqrt{2\pi}\theta_{-,s}),
\end{equation}
where, at weak coupling, the interaction strength $g$ is proportional to $J_\perp / (2\pi^2 \alpha^2)$, with $\alpha$ being the short-distance cutoff (see SM). As shown in the SM, this term is renormalization-group (RG) relevant and flows to strong coupling. In the strong-coupling limit ($g \to \infty$), the cosine term pins the fields to values such as $(\theta_{-,s}, \phi_{+,s}) = (0,0)$ or $(\pi/2, \pi/2)$ up to a factor of $\sqrt{2\pi}$. Consequently, both spin modes become gapped in the low-energy limit, leaving only two gapless charge modes. This is consistent with our DMRG observations of a finite spin gap [Fig.~\ref{fig:2}(a)] and a central charge $c=2$ [Fig.~\ref{fig:4}(b)].

To identify the dominant orders, we evaluate the long-range correlation of physical operators.  
We found the interlayer PDW operator takes the form  
\begin{equation}
\mathcal{O}_{\mathrm{PDW},\perp} \sim 
e^{i(\pm \phi_{-,c} - \theta_{+,c})},
\end{equation}
which exhibits quasi-long-range correlations with $q=\delta k_F$. 
We further analyze the composite intralayer PDW operator  
$ \mathcal{O}_{\mathrm{PDW},\parallel} ^{(2)}\sim \vec{N}^{(1)} \cdot \vec{\Delta}^{(2)}_T$,  
where $\vec{N}^{(1)}$ is the $2k_{F,1}$ SDW on leg~1 which continuously connects to N\'eel order at $P=1$, 
and $\vec{\Delta}^{(2)}_T$ denotes triplet pairing on leg~2.  
Note that this composite order has the same wave-vector of intralayer PDW correlation $\Delta_{\parallel,2}$ observed in DMRG. 
This order is bosonized as, 
\begin{equation}
\mathcal{O}_{\mathrm{PDW},\parallel}^{(2)}
\sim
e^{i (\theta_{+,c}-\theta_{-,c}-\phi_{+,c}-\phi_{-,c})},
\end{equation}
which again shows quasi-long-range correlation with $q=2k_{F,1}$.
The bosonized forms of $\mathcal{O}_{\mathrm{PDW},\parallel}^{(1)}$ only differ by a sign factor in the exponential from $\mathcal{O}_{\mathrm{PDW},\parallel}^{(2)}$ 
(see SM).
In Table~\ref{table:1}, we summarize and compare the power-law exponents of those variables. 
We show the interlayer PDW correlations decay more slowly than the composite intralayer PDW channel, establishing the interlayer iC-PDW as the dominant instability of the $c=2$ phase. This is the consistent with the DMRG results [Figs.~\ref{fig:2}(c,d)].

\begin{table}[tb]
    \centering
\begin{tabular}{c|c|c|c}
\hline
\hline
Order &  Definition & Momentum & Exponent  $\alpha$ \\ \hline
$\mathcal{O}_{\mathrm{PDW},\perp}$ 
&  $\epsilon_{\sigma,\sigma'}\, c_{1,\sigma}^\dagger c_{2,\sigma'}^\dagger$
& $\delta k_{F}$
& $\dfrac{1}{2}\Big[K_{-,c}+\dfrac{1}{K_{+,c}}\Big]$\\[2pt]
$\mathcal{O}_{\mathrm{PDW},\parallel}^{(2)}$ 
& $\vec{N}^{(1)} \cdot \vec{\Delta}^{(2)}_T$
& $2k_{F,1}$
& $\dfrac{1}{2}\displaystyle\sum_{\sigma =\pm }\Big[ K_{\sigma,c}+\dfrac{1}{K_{\sigma,c}}\Big]$   \\
\hline
\hline
\end{tabular}
\caption{PDW correlations in bosonization.  
$\mathcal{O}_{\mathrm{PDW},\perp}$ and $\mathcal{O}_{\mathrm{PDW},\parallel}$ are the interlayer
and intralayer (composite) PDW, respectively.
$\alpha_\perp, \alpha_\parallel$ is power-law exponent of pair correlator,
$\langle \Delta^\dagger(0)\Delta(r)\rangle \sim |r|^{-\alpha}$.
We find $\alpha_\perp < \alpha_\parallel$.
}
\label{table:1}
\end{table}

\begin{figure}[tb]
    \centering
\includegraphics[width=\linewidth]{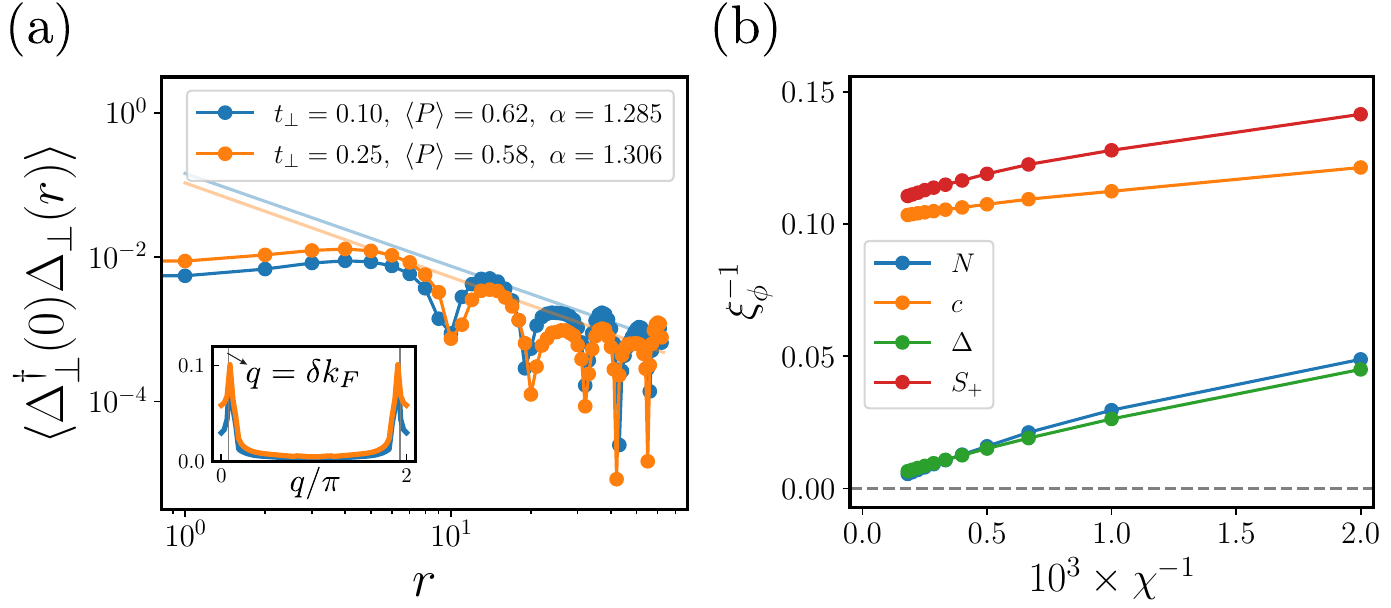}
    \caption{DMRG results of interlayer iC-PDW in the presence of $t_\perp$. We use $t=J_\perp=1$ and $J=2$ with $n=7/8$ and add potential difference between two layers $\epsilon=0.6$ for a finite polarization $\langle P \rangle$.
(a) Real-space pair correlation functions for $t_\perp = 0.1$ and $0.25$ ($L_x=64, \chi=4500$). The pair correlations exhibit power-law decay, indicating robust quasi-long-range order upon finite $t_\perp$. Inset shows its Fourier transform, $\langle \Delta_\perp ^\dagger (q) \Delta_\perp (-q)\rangle$, which shows a peak at $q=\delta k_F =(1-n)P \pi$. 
(b) Correlation lengths of different operators at $t_\perp = 0.25$ ($L_x=16$, $\chi\leq 5500$). 
The corresponding charge sectors are specified in terms of $(\delta N_\mathrm{tot},2\delta S_z)$ as 
$(0,0)$, $(1,1)$, $(2,0)$, $(0,2)$, for 
$N$, $c$, $\Delta$, and $S_+$, respectively.
}
    \label{fig:5}
\end{figure}

\paragraph{Effect of interlayer hopping—} Having established the interlayer iC-PDW phase in $t$--$J$--$J_\perp$ model, we now consider the inclusion of a finite interlayer hopping $t_\perp$. The inclusion of $t_\perp$ breaks the conservation of individual leg particle numbers, leaving only the total particle number $N = N_1 + N_2$ conserved. To maintain a finite polarization and explicitly break mirror symmetry, we introduce a layer-dependent potential difference,
$\epsilon \sum_i (n_{i,2} - n_{i,1})$.

From bosonization, a second order process of the $t_\perp$ term may lead to an instability of the $c=2$ PDW phase towards a $c=1$ phase with only charge-4e pairing correlations (See SM Sec.~\textcolor{red}{II} for details). In this phase, the charge-$2e$ PDW order is gapped. However, this gap remains exceedingly small and is nearly undetectable in our finite-size numerical simulations ($L=64$, $\chi=5500$). With the system size reachable by our current DMRG calculation, the interlayer pairing correlations continue to exhibit power-law decay with a characteristic wave vector $q = \delta k_F$, even with $t_\perp$~[Fig.~\ref{fig:5}(a)]. Furthermore, iDMRG calculations show that the correlation length associated with charge-$2e$ pairing still increases with increasing bond dimension~[Fig.~\ref{fig:5}(b)]. However, in SM, we show that the central charge $c$ fit for $t_\perp=0.25$ has a tendency to decrease from $2$ towards $1$ when increasing the bond dimension~[Fig.~\ref{fig:s6}]. 
We conjecture that the phase eventually flow to the $c=1$ charge-4e pairing phase in the infinite correlation length limit beyond our computation ability. Nevertheless, the PDW order still have a correlation length which is at least $\xi_{\Delta}=152.52$ approaching to $\chi=5500$~[Fig.~\ref{fig:5}(b)] and the phase may not be distinguished with the PDW phase at $t_\perp=0$ in experiments within this large length scale.

\textit{Discussion.---}
Beyond the primary focus of this work, we briefly comment on the spin-gapless regime at small $J$ shown in Fig.~\ref{fig:1}(a).
This regime exhibits power-law interlayer pairing correlations with a central charge $c=3$ (see SM).
A natural conjecture is that this phase arises from gapping out one degree of freedom starting from a $c=4$ Luttinger liquid. However, at present, we do not have a consistent theoretical framework that simultaneously explains both the PDW correlations and the emergence of $c=3$. 
We leave a detailed understanding of this regime and its phase transition for future work.

In conclusion, we have shown that the two-leg $t$--$J$--$J_\perp$ model offers a new route to realizing interlayer incommensurate PDW correlations via particle-number imbalance. When the particle-number of each leg is conserved, this phase is robust. 
However, a small interlayer tunneling $t_\perp$ induces an instability toward a phase with zero-momentum charge-4e correlations.
An important open question is whether such PDW correlations can persist beyond one dimension. Exploring higher-dimensional generalizations or $2N$-leg extensions of the $t$--$J$--$J_\perp$ model may provide a bridge to previous weak-coupling studies in the context of bilayer nickelate~\cite{fan2025minimal,oh2025pairdensitywave}, and we leave this direction for future work. 
Finally, we emphasize that the $t$--$J$--$J_\perp$ ladder can be engineered in cold-atom quantum simulators by applying an interlayer potential offset~\cite{bohrdt2022strong,hirthe2023magnetically}. Given the tunability of polarization in these systems, our theory propose future experiments targeting the finite-polarization regime to probe the polarization-driven PDW and phase transitions.

\textit{Acknowledgements ---} We thank Y.-M. Wu for useful discussions. 
H.O. and Y-H.Z. are supported by a startup fund from Johns Hopkins University and the Alfred P. Sloan Foundation
through a Sloan Research Fellowship (YHZ).
J.M.M. is supported by a startup fund at Stanford University.  

%

\onecolumngrid
\newpage
\clearpage
\setcounter{equation}{0}
\setcounter{figure}{0}
\setcounter{table}{0}
\setcounter{page}{1}
\setcounter{section}{0}

\maketitle 
\makeatletter
\renewcommand{\theequation}{S\arabic{equation}}
\renewcommand{\thefigure}{S\arabic{figure}}
\renewcommand{\thetable}{S\arabic{table}}
\renewcommand{\thesection}{S\arabic{section}}

\appendix
\onecolumngrid

\begin{center}
\vspace{10pt}
\textbf{\large Supplemental Material for ``Incommensurate pair-density-wave correlations in two-leg ladder $t$--$J$--$J_\perp$ model''}
\end{center} 
\begin{center}
{Hanbit Oh$^{1}$, Julian May-Mann$^{2}$, Ya-Hui Zhang$^{1}$}\\
\emph{$^{1}$ William H. Miller III Department of Physics and Astronomy, \\
Johns Hopkins University, Baltimore, Maryland, 21218, USA}\\
\emph{$^{2}$ Department of Physics, Stanford University, Stanford, CA 94305, USA}
\vspace{5pt}
\end{center}
\tableofcontents

\section{I. Details on DMRG Simulation}

\subsection{A. Doping and polarization dependence}
\begin{figure}[h]
    \centering
\includegraphics[width=0.4\linewidth]{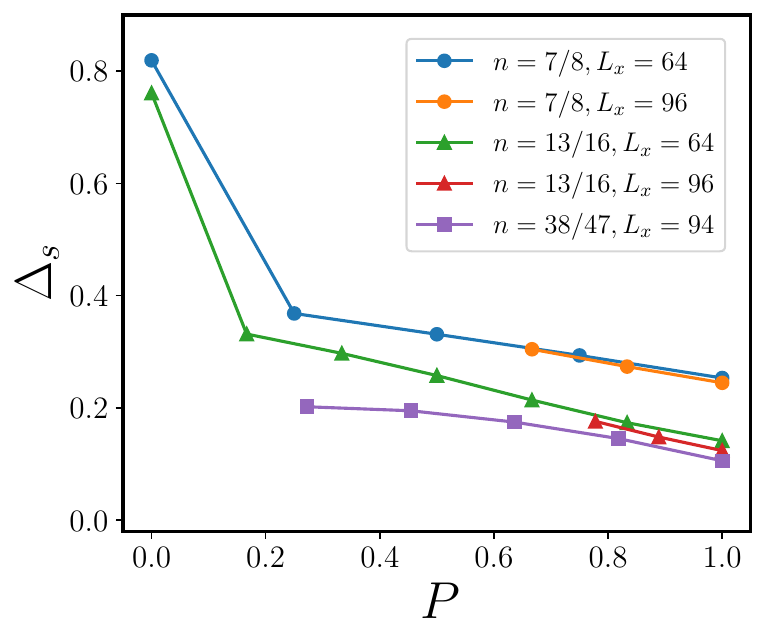}
    \caption{Doping and polarization dependence on spin-gap for $t=1$ and $J_\perp = J = 2$ with $\chi=2500$.
Results are shown for fillings $n=7/8$, $13/16$, and $38/47$.
A finite spin gap persists for all $0 \le P \le 1$ and increases with electron density.    }
    \label{fig:s1}
\end{figure}

In the main text, we present results for (i) $n = 7/8$ with $P = m/4$ ($m = 0,\dots,4$). 
To confirm the robustness of the $c=2$ PDW phase over a broad parameter range and to check convergence, in Fig.~\ref{fig:s1}, we additionally show data for 
(ii) $n = 13/16=0.8125$ with $P = m/6$ ($m = 0,\dots,6$), and 
(iii) $n = 38/47 \simeq 0.8085$ with $P = m/11$ ($m = 0,\dots,11$).
Since $\delta k_F = (1-n) P \pi$, case (iii) lies extremely close to the incommensurate limit.  
For convergence with respect to system size, we use $L_x = 64$ and $96$ for cases (i) and (ii).

\subsection{B. More DMRG results for spin-gapless regime $J<J_c$}
Here, we provide more DMRG calculation results for spin-galess region. 
For $J<J_c$, we find that the spin gap closes; however, the interlayer pairing correlations 
still exhibit a slowly decaying power law.  
A detailed characterization of such phases and the nature of the spin-gap transition 
is left for future work.

\begin{figure}[h]
    \centering
\includegraphics[width=.85\linewidth]{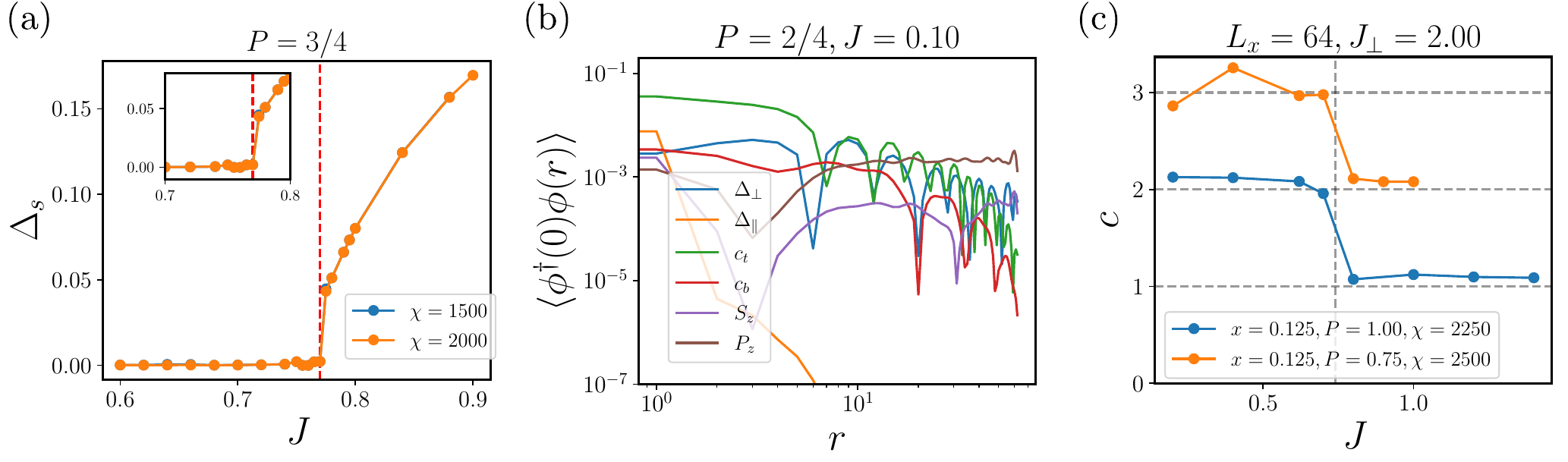}
\caption{
fDMRG results at $n = 7/8$, $t = 1$, and $J_\perp = 2$ with bond dimension $\chi = 2000$.  
(a) Zoom-in of the spin-gap transition region at $P = 3/4$.  The spin gap discontinuosly closes for $J \lesssim 0.78$.  
(b) Various-correlation functions at $J = 0.1$. Interlayer pair-correlation shows a power-law decay despite the absence of a spin gap.
(c) Central charge extracted from iDMRG.
At partial polarization $P=3/4$, the central charge transitions from $c \simeq 2$ for $J < 0.78$ (spin-gapped region)
to $c \simeq 3$ for $J > 0.78$ (spin-gapless region).
}

    \label{fig:s2}
\end{figure}

\subsection{C. More DMRG results for spin-gapped regime $J>J_c$}
We present additional DMRG results for the spin-gapped region $J > J_c$.
In the main text, we only show the Fourier transform of the intralayer pairing correlations on leg~2,
which is the less populated leg.
Here we show that the intralayer pairing correlations on leg~1 do not exhibit a sharp finite-momentum peak,
in contrast to those on leg~2, which display a clear peak at $2k_{F,1}$.

\begin{center}
    \begin{figure}[h]
        \centering
\includegraphics[width=0.4\linewidth]{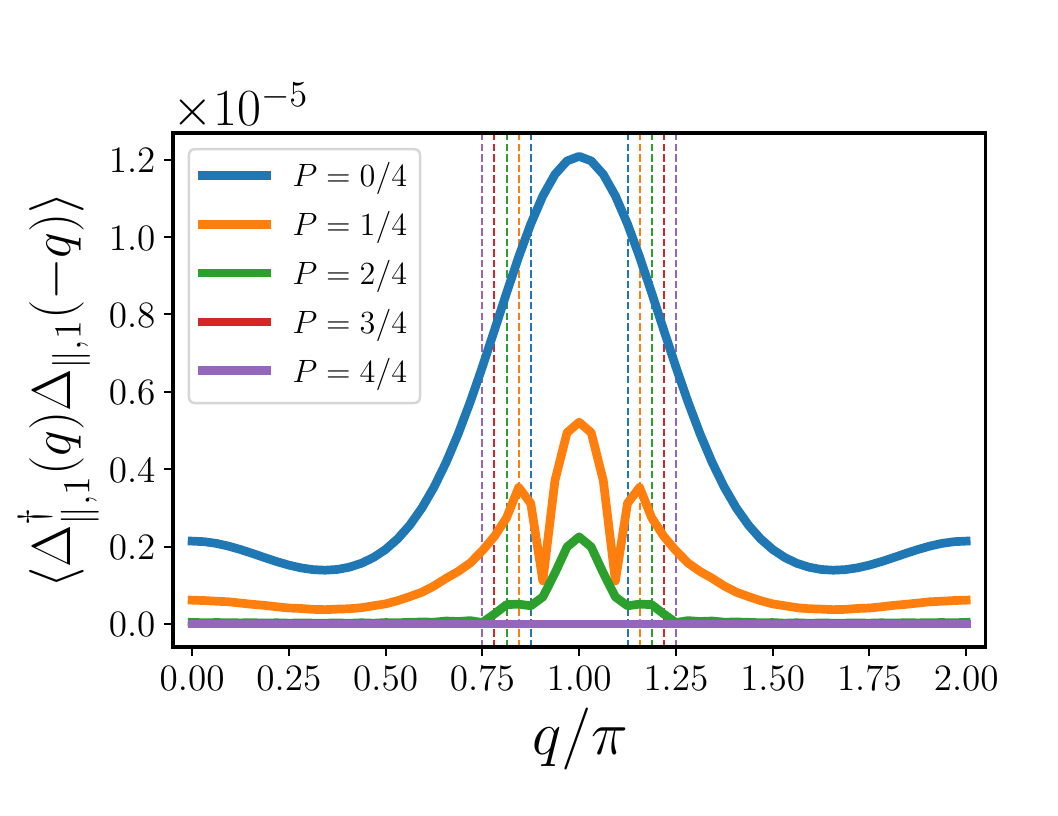}
\caption{Fourier transform of the intralayer pairing correlations on layer~1.
We use a filling of $n = 7/8$ with parameters $t = 1$ and $J_\perp = J = 2$.
No sharp peak of $\Delta_{\parallel,1}$ is observed for any value of $P$.
The dashed line indicates $q = 2k_{F,2}$.
}
        \label{fig:s3}
    \end{figure}
\end{center}

\begin{figure}[h]
    \centering
\includegraphics[width=0.6\linewidth]{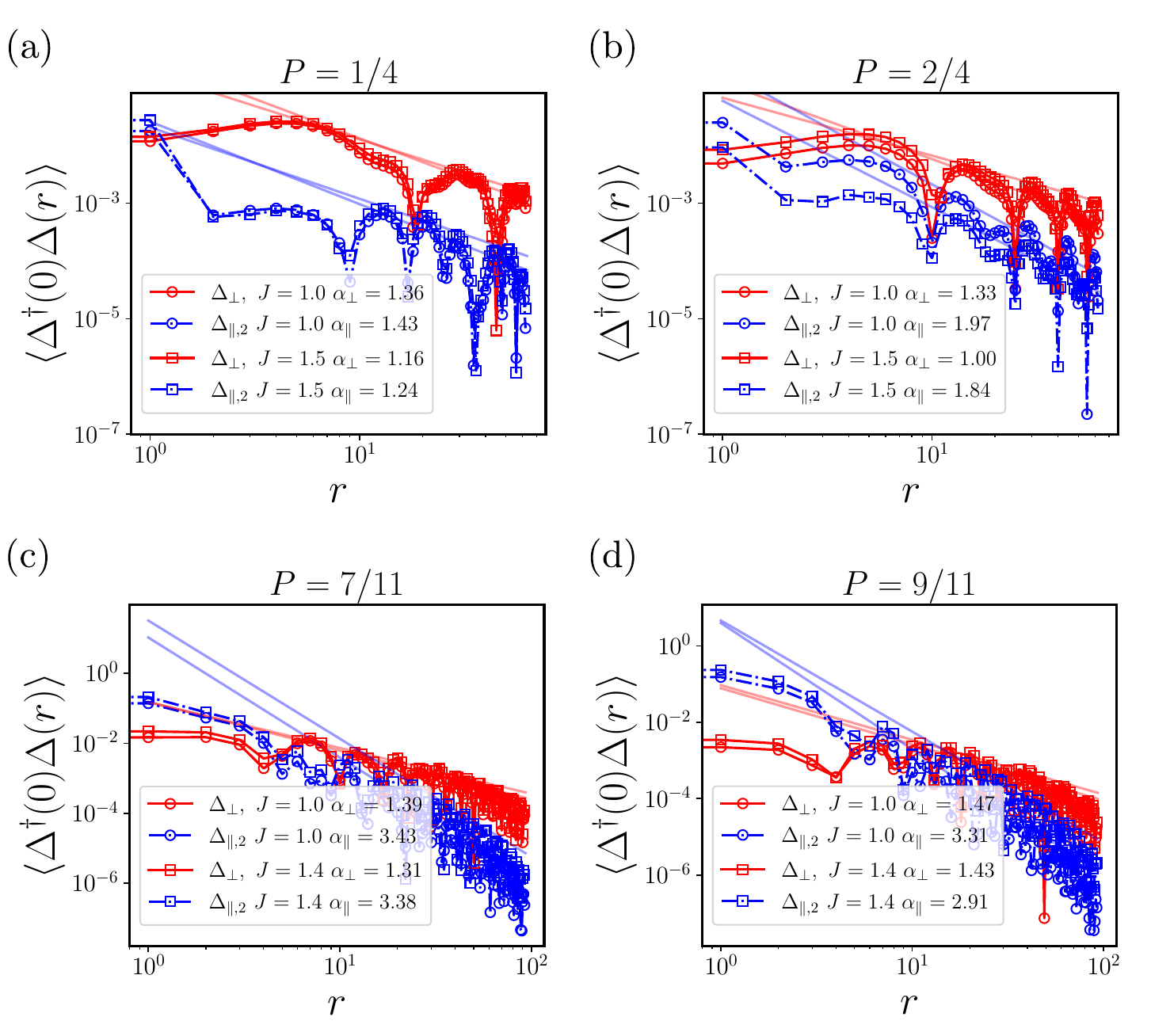}
\vspace{-10pt}
    \caption{Pair correlation function from fDMRG $t=1$, $J_\perp = 2$, with $\chi=2000$.  
    (a,b) are at $n=7/8$ and (c,d) are at $n=38/47$.} 
    \label{fig:s4}
\end{figure}

\begin{figure}[h]
    \centering
\includegraphics[width=.9\linewidth]{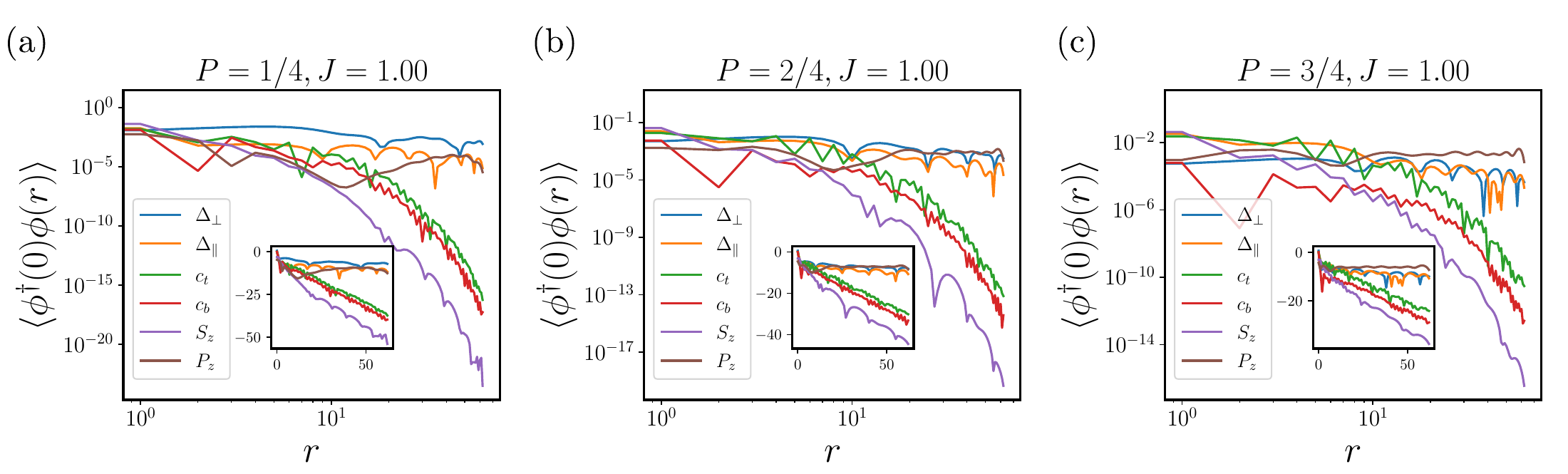}
    \caption{Correlation functions of various operators obtained from fDMRG at filling
$n=7/8$ with parameters $t=1$, $J_\perp=2$, and $J=1$, using a bond dimension
$\chi=2000$. The main panel shows the data on a log--log scale, while the inset
presents the same correlations on a linear--log scale.
 }
    \label{fig:s5}
\end{figure}

In Fig.~\ref{fig:s4}, we show the pair-correlation data for various dopings and polarizations, demonstrating that the interlayer PDW correlation is the dominant one. 
We further evaluate other correlation functions. In the broad parameter range, we found
\begin{eqnarray}
    \Delta_\perp, \Delta_\parallel, P_z: && \ \text{Gapless}, \\
   c_t,c_b,S_z: &&\  \text{Gapped}.
\end{eqnarray}
as illustrated in Fig.~\ref{fig:s5}.

\subsection{D. Effect of interlayer hopping $t_\perp$}
\begin{figure}[h]
    \centering
\includegraphics[width=0.65\linewidth]{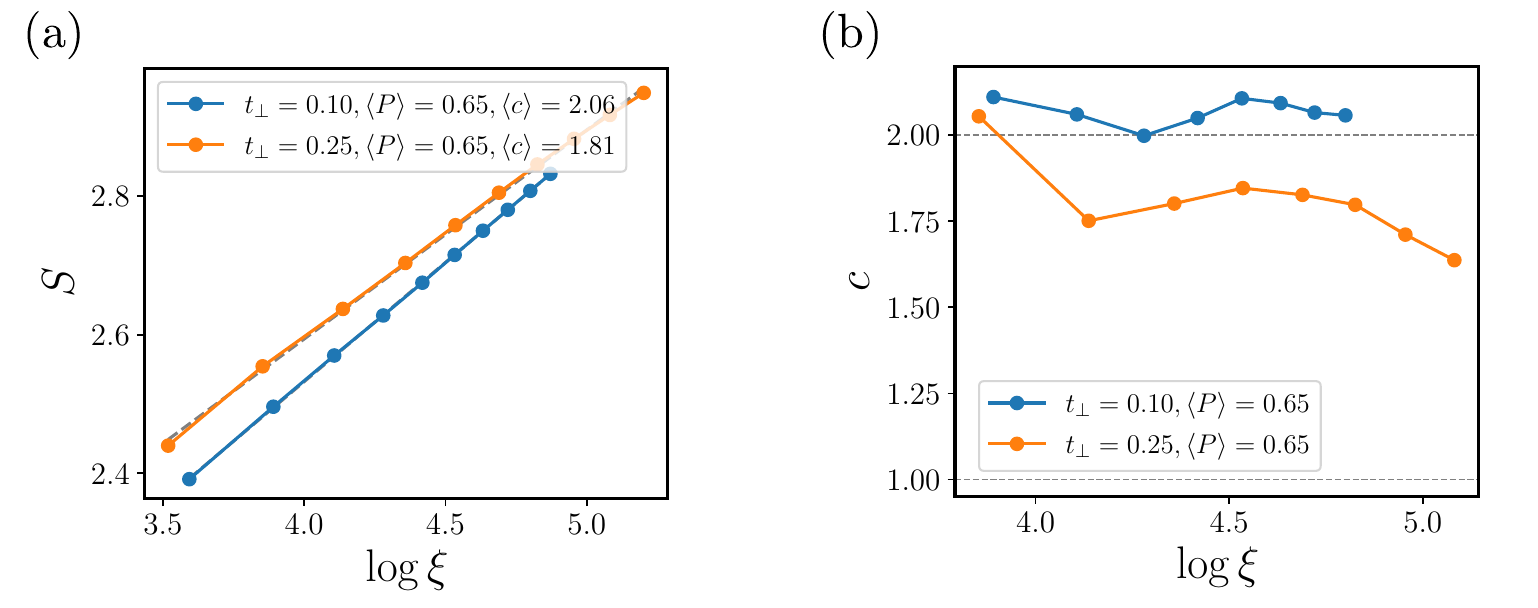}
    \caption{iDMRG results at filling $n=7/8$ with $t=1$ and $J=J_\perp=2$. A layer-dependent potential $\epsilon=0.6$ is applied to induce a finite polarization $\langle P\rangle\neq 0$. The simulations are performed with bond dimensions up to $\chi\le 5500$. 
(a) Entanglement entropy $S$ as a function of correlation length $\xi$. 
(b) Central charge $c$ extracted from the scaling relation
$c = 6\,\partial S / \partial(\log \xi)$ as a function of $\xi$. 
For $t_\perp=0.1$ and $0.25$, we find $c\simeq 2$ over a broad range of correlation lengths. 
For larger $t_\perp$, however, the extracted central charge shows a tendency to deviate from $c=2$ and approach $c=1$ at larger bond dimensions or longer correlation lengths, suggesting a possibility to flow toward a $c=1$ phase at sufficiently large $t_\perp$.
}
    \label{fig:s6}
\end{figure}

In the main text, we show that the iC-PDW phase stable under $t_\perp$ term. In Fig~\ref{fig:s6}, we present entanglement entropy plot to extract the central charge. 

\section{II. Details of the Bosonization analysis}

Following the convention of Ref.~\cite{giamarchi2003quantum}, the electron
operator is represented as
\begin{eqnarray}
    c_{\ell,\sigma}(x) 
    = R_{\ell,\sigma}(x)\, e^{i k_{F,\ell}x}
    + L_{\ell,\sigma}(x)\, e^{-i k_{F,\ell}x},
\end{eqnarray}
with
\begin{eqnarray}
    r_{\ell,\sigma}(x)
    &=& \frac{1}{2\pi \alpha}\,
    U_{r,\ell,\sigma}\,
    \exp\!\Bigg[
        -\frac{i}{\sqrt{2}}
        \Big(
            r\phi_{\ell,c} - \theta_{\ell,c}
            + \sigma ( r \phi_{\ell,s} - \theta_{\ell,s} )
        \Big)
    \Bigg],
\end{eqnarray}
where $r = R,L$ labels right- and left-moving modes, $\ell=1,2$ is the
layer index, and $\sigma=\uparrow,\downarrow$ is the spin index.
The fields $\theta_{\ell,\rho}$ and $\phi_{\ell,\rho}$ ($\rho=c,s$) are real
bosonic fields describing charge and spin degrees of freedom, respectively:
\begin{eqnarray}
\phi_{\ell,c}&=&
\frac{1}{\sqrt{2}}\big(\phi_{\ell,\uparrow}+ \phi_{\ell,\downarrow}\big),
\quad 
\phi_{\ell,s}=
\frac{1}{\sqrt{2}}\big(\phi_{\ell,\uparrow}-\phi_{\ell,\downarrow}\big),
\\
\theta_{\ell,c}&=&
\frac{1}{\sqrt{2}}\big(\theta_{\ell,\uparrow}+\theta_{\ell,\downarrow}\big),
\quad
\theta_{\ell,s}=
\frac{1}{\sqrt{2}}\big(\theta_{\ell,\uparrow}-\theta_{\ell,\downarrow}\big),
\end{eqnarray}
with the standard commutation relation
\begin{equation}
    \big[\phi_{\ell,\rho}(x), \partial_y\theta_{\ell',\rho'}(y)\big]
    = i \pi\, \delta_{\ell,\ell'}\, \delta_{\rho,\rho'}\, \delta (x-y).
\end{equation}
Here $U_{r,\ell,\sigma}$ are Klein factors ensuring the correct
anticommutation relations between fermions.

At $J_\perp=0$, the Hamiltonian density of the two-leg
$t$–$J$–$J_\perp$ model reduces to
\begin{align}
\mathcal{H}_0 
&= \sum_{\ell=1,2}
\left\{
    \frac{v_{\ell,c}}{2\pi}
    \Big[
        K_{\ell,c}(\partial_x \theta_{\ell,c})^2 
        + \frac{1}{K_{\ell,c}}(\partial_x \phi_{\ell,c})^2
    \Big]
    +
    \frac{v_{\ell,s}}{2\pi}
    \Big[
        (\partial_x \theta_{\ell,s})^2 + (\partial_x \phi_{\ell,s})^2
    \Big]
\right\}.
\end{align}
Here $K_{\ell,c}$ is the charge Luttinger parameter determined by $t$, $J$,
and the filling $n_\ell$, while $v_{\ell,c}$ and $v_{\ell,s}$ denote the
charge and spin velocities, respectively.

To express the interlayer spin interaction, we first bosonize the spin
operators on each leg:
\begin{eqnarray}
    S_{\ell}^z (x) &=& 
    -\frac{1}{\sqrt{2}\pi} \partial_x \phi_{\ell,s}
    +\frac{1}{2\pi \alpha}
    \sin\!\big(\sqrt{2}\,\phi_{\ell,s}\big)
    \Big[
        e^{-2ik_{F,\ell}x}e^{i \sqrt{2}\phi_{\ell,c}}
        +e^{2ik_{F,\ell}x}e^{-i \sqrt{2}\phi_{\ell,c}}
    \Big],
    \label{eq:s_cz}
    \\
       S_{\ell}^+ (x) &=& 
\frac{e^{-i \sqrt{2}\theta_{\ell,s}}}{2\pi \alpha}
\Big\{
    i\Big[
        e^{-2ik_{F,\ell}x}
        e^{i \sqrt{2}\phi_{\ell,c}}
        +e^{2ik_{F,\ell}x}
        e^{-i \sqrt{2}\phi_{\ell,c}}
    \Big]
    +2 \sin\!\big(\sqrt{2}\,\phi_{\ell,s}\big)
\Big\},
        \label{eq:s_c+}
\end{eqnarray}
where we follow the choice of Klein factors in Ref.~\cite{Zhang2022Pair}.
The interlayer spin interaction is then written as
\begin{eqnarray}
    \mathcal{H}_\perp &=& g
    \Big[
        S^z_{1}(x)S^z_{2}(x)
        +\frac{1}{2} \Big( S^+_{1}(x)S^-_{2}(x)
        +S^-_{1}(x) S^+_{2}(x)
        \Big)
    \Big]
\\
     &\simeq& g
\left[
   \frac{1}{2\pi^2}
   \partial_x \phi_{1,s}\,
   \partial_x \phi_{2,s}
   +\frac{1}{4\pi^2 \alpha^2}
   \sin\!\big(\sqrt{2}\,\phi_{1,s}\big)
   \sin\!\big(\sqrt{2}\,\phi_{2,s}\big)
   \cos\!\big(\sqrt{2}(\theta_{1,s}-\theta_{2,s})\big)
\right]
\\
    &=& g
\left[
   \frac{1}{4\pi^2}
   \big((\partial_x \phi_{+,s})^2- (\partial_x \phi_{-,s})^2\big)
   -\frac{1}{2\pi^2 \alpha^2}
   \big(\cos 2 \phi_{+,s}-\cos 2\phi_{-,s}\big)\cos 2\theta_{-,s}
\right],
\end{eqnarray}
where we have kept only the slowly varying contributions and set $g=J_\perp$.

The coupling $g$ renormalizes the quadratic part of the Hamiltonian.  
In the following, we assume that the bare differences between the two legs
are not substantial, so that $K_{\ell,c} = K_{c}$ and $v_{\ell,c} = v_{c}$
for both layers.  
The full Hamiltonian is then conveniently written in the
$\{\phi_{\pm,c/s},\, \theta_{\pm,c/s}\}$ basis as
\begin{eqnarray}
\mathcal{H} 
&=& \frac{v_{c}}{2\pi}
\Big[
K_{c}\left(\partial_x \theta_{+,c}\right)^2 
+ \frac{1}{K_{c}}\left(\partial_x \phi_{+,c}\right)^2 
\Big]
+
\frac{v_{c}}{2\pi}
\Big[
K_{c}\left(\partial_x \theta_{-,c}\right)^2 
+ \frac{1}{K_{c}}\left(\partial_x \phi_{-,c}\right)^2 
\Big]
\nonumber \\
&&+
\frac{v_{+,s}}{2\pi}
\Big[
K_{+,s}\left(\partial_x \theta_{+,s}\right)^2 
+ \frac{1}{K_{+,s}}\left(\partial_x \phi_{+,s}\right)^2 
\Big]
+
\frac{v_{-,s}}{2\pi}
\Big[
K_{-,s}\left(\partial_x \theta_{-,s}\right)^2 
+ \frac{1}{K_{-,s}}\left(\partial_x \phi_{-,s}\right)^2 
\Big]
\nonumber \\
&&
-\frac{g}{2\pi^2 \alpha^2}
\cos\!\left(2\phi_{+,s}\right)
\cos\!\left(2\theta_{-,s}\right),
\label{eq:bosonized_ham}
\end{eqnarray}
where the renormalized spin-sector parameters are given by
\begin{eqnarray}
v_{\pm,s}
&=& 
v_s \sqrt{1\pm \frac{g}{2\pi v_s}}, 
\quad 
K_{\pm,s}
= \frac{1}{\sqrt{1\pm \frac{g}{2\pi v_s}}}.
\end{eqnarray}
In Eq.~\ref{eq:bosonized_ham} we have discarded the term 
$\cos 2\phi_{-,s}\,\cos 2\theta_{-,s}$, whose coupling has scaling
dimension $[g'] = 2 - K_{-,s} + \frac{1}{K_{-,s}} < 0$ and is therefore
irrelevant in the RG sense.  
This vertex does not affect the low-energy physics and can be safely
neglected.

The one-loop RG equation of the remaining couplings are given by 
\begin{eqnarray}
    \frac{d K_{+,s}}{dl}
    &=& -\frac{1}{4} K_{+,s}^2\, g^2,
    \\
    \frac{d K_{-,s}}{dl}
    &=& \frac{1}{4} g^2,
    \\
    \frac{dg}{dl}
    &=& \left[ 2 - K_{+,s} - \frac{1}{K_{-,s}} \right] g.
\end{eqnarray}
For any initial value $g(0) > 0$, the coupling flows to $g \to +\infty$.
At this strong-coupling fixed point, the spin fields $\phi_{+,s}$ and
$\theta_{-,s}$ are pinned, opening a finite spin gap.  
Consequently, only the charge sector remains gapless, and the low-energy
theory has central charge $c=2$.

\subsection{A. Properties of order parameters}

We now analyze the long-distance behavior of various order parameters at the
strong-coupling fixed point ($g\to\infty$).  
At this point, $\phi_{+,s}$ and $\theta_{-,s}$ are pinned, whereas their
dual fields $\theta_{+,s}$ and $\phi_{-,s}$ are disordered.  

First, consider the interlayer PDW operator,
\begin{eqnarray}
 \mathcal{O}_\mathrm{PDW,\perp}
&=&
    c_{1,\uparrow}\,
    c_{2,\downarrow}
  - c_{1,\downarrow}\,
    c_{2,\uparrow}.
\end{eqnarray}
The $q=\delta k_F$ component of this operator can be decomposed as
\begin{eqnarray}
    R_{1,\uparrow}
    L_{2,\downarrow}
    &=&  \frac{1}{2\pi \alpha}\,
    e^{-i(\phi_{-,c}-\theta _{+,c} +\phi_{+,s} -\theta_{-,s})}
    =\frac{1}{2\pi \alpha}\,
    e^{i(-\phi_{-,c}+\theta _{+,c})},
\\
   R_{1,\downarrow}
   L_{2,\uparrow}
    &=&  \frac{1}{2\pi \alpha}\,
    e^{-i(\phi_{-,c}-\theta _{+,c} -\phi_{+,s} +\theta_{-,s})}
    =\frac{1}{2\pi \alpha}\,
    e^{i(-\phi_{-,c}+\theta _{+,c})},
\\
    L_{1,\uparrow}
    R_{2,\downarrow}
    &=&  \frac{1}{2\pi \alpha}\,
    e^{+i(\phi_{-,c}+\theta _{+,c} +\phi_{+,s} +\theta_{-,s})}
    =\frac{1}{2\pi \alpha}\,
    e^{i(\phi_{-,c}+\theta _{+,c})},
\\
    L_{1,\downarrow}
   R_{2,\uparrow}
    &=&  \frac{1}{2\pi \alpha}\,
    e^{+i(\phi_{-,c}+\theta _{+,c} -\phi_{+,s} -\theta_{-,s})}
    =\frac{1}{2\pi \alpha}\,
    e^{i(\phi_{-,c}+\theta _{+,c})}.
\end{eqnarray}
Hence the interlayer PDW operator is expressed purely in terms of the
gapless charge fields,
\begin{eqnarray}
\mathcal{O}_\mathrm{PDW,\perp} \sim 
    e^{i(\pm\phi_{-,c}+\theta _{+,c})},
    \label{eq:pdw_inter_boson}
\end{eqnarray}
and therefore exhibits quasi-long-range order.  
The corresponding two-point function decays as
\begin{eqnarray}
 \big\langle
 \mathcal{O}^\dagger_\mathrm{PDW,\perp}(0)\,
 \mathcal{O}_\mathrm{PDW,\perp} (r)
 \big\rangle
 \sim |r|^{-\alpha_\perp},
\end{eqnarray}
with exponent
\begin{eqnarray}
 \alpha_\perp = \frac{1}{2}\left[K_{-,c}+\frac{1}{K_{+,c}}\right].
\end{eqnarray}

We next consider the intralayer composite PDW order of layer~2,
\begin{eqnarray}
\mathcal{O}_{\mathrm{PDW},\parallel}^{(2)}
   &\sim& \vec{N}^{(1)} \cdot \vec{\Delta}^{(2)}_T.
\end{eqnarray}
Here, $\vec{\Delta}^{(2)}_T$ denotes the triplet pairing operator on leg~2,
and $\vec{N}^{(1)}$ is the $2k_{F,1}$ component of the spin operator on
leg~1 [Eqs.~\ref{eq:s_cz}–\ref{eq:s_c+}], which continuously connects
to the N\'eel order at $P=1$.
One finds $\vec{\Delta}_T^{(2)}\sim e^{i\sqrt{2}\theta_{2,c}}
(\sin\sqrt{2}\theta_{2,s},\cos\sqrt{2}\theta_{2,s},\sin \sqrt{2}\phi_{2,s})$ and
\begin{eqnarray}
   N_{+}^{(1)}
    &=&     \frac{i }{2\pi \alpha}\,
    e^{-i \sqrt{2}\theta_{1,s}}
    e^{-i \sqrt{2}\phi_{1,c}}
    =N_{x}^{(1)}+iN_{y}^{(1)},\\
    N_{-}^{(1)}
    &=&     \frac{-i}{2\pi \alpha}\,
    e^{+i \sqrt{2}\theta_{1,s}}
    e^{-i \sqrt{2}\phi_{1,c}}
    =N_{x}^{(1)}-iN_{y}^{(1)},\\
    N_{z}^{(1)}
    &=& 
    \frac{1}{2\pi \alpha}
    \sin \sqrt{2} \phi_{1,s}\,
    e^{-i \sqrt{2}\phi_{1,c}},
\end{eqnarray}
so that
$\vec{N}^{(1)}\sim e^{-i\sqrt{2}\phi_{1,c}}
(\sin \sqrt{2}\theta_{1,s},\cos \sqrt{2}\theta_{1,s}, \sin \sqrt{2} \phi_{1,s})$.
The composite order at layer~2 then bosonizes as
\begin{equation}
\mathcal{O}_{\mathrm{PDW},\parallel}^{(2)}\sim
e^{i\sqrt{2} \theta_{2,c}}\, e^{-i\sqrt{2}\phi_{1,c}}
\sim
e^{i (\theta_{+,c}-\theta_{-,c}-\phi_{+,c}-\phi_{-,c})},
\label{eq:pdw_intra_boson}
\end{equation}
which again involves only the gapless charge fields and thus exhibits
quasi-long-range correlations.  

Similarly, we find that the bosonized form of
\begin{eqnarray}
\mathcal{O}_{\mathrm{PDW},\parallel}^{(1)}
   &\sim& \vec{N}^{(2)} \cdot \vec{\Delta}^{(1)}_T
\end{eqnarray}
is given by
\begin{equation}
\mathcal{O}_{\mathrm{PDW},\parallel}^{(1)} \sim
e^{i\sqrt{2}\theta_{1,c}}\, e^{-i\sqrt{2}\phi_{2,c}}
\sim
e^{i(\theta_{+,c} + \theta_{-,c} - \phi_{+,c} + \phi_{-,c})}.
\label{eq:pdw_intra_boson_1}
\end{equation}
The bosonized form of $\mathcal{O}_{\mathrm{PDW},\parallel}^{(1)}$
differs from that of $\mathcal{O}_{\mathrm{PDW},\parallel}^{(2)}$
only by a sign factor in the exponential.

The corresponding two-point function decays as
\begin{eqnarray}
 \big\langle
\mathcal{O}^{\dagger}_\mathrm{PDW,\parallel}(0)\, \mathcal{O}_\mathrm{PDW,\parallel} (r)
 \big\rangle
 \sim |r|^{-\alpha_\parallel},
 \quad
 \ell=1,2,
\end{eqnarray}
with exponent
\begin{eqnarray}
    \alpha_\parallel
    = \frac{1}{2}\left[
        K_{+,c}+K_{-,c}
        +\frac{1}{K_{+,c}}+\frac{1}{K_{-,c}}
    \right].
\end{eqnarray}
From the above analysis, the scaling dimensions of the intralayer PDW orders in the two layers are found to be the same.
This conclusion is based on the simple assumption that the Luttinger parameters of each layer are the same, $K_1 = K_2$.
In general, however, it is very likely that $K_1 \neq K_2$ in realistic situations,
and the scaling dimensions of the two intralayer components can therefore be different, as demonstrated in our DMRG results.

\subsection{B. Effect of interlayer hopping $t_\perp$}

We now briefly comment on the effect of a small interlayer hopping in the UV theory
\begin{equation}
    H_{t_\perp}
    = - t_\perp \sum_{i,\sigma}
    \left(
        c^{\dagger}_{i1\sigma} c_{i2\sigma}
        + \text{H.c.}
    \right),
\end{equation}
which leads to the continuum Hamiltonian
\begin{eqnarray}
    \mathcal{H}_{t_\perp}
    &\sim&
    - t_\perp \sum_{\sigma}
    \Big[
        R^{\dagger}_{1\sigma} R_{2\sigma}\, e^{-i\delta k_F x}
        + L^{\dagger}_{1\sigma} L_{2\sigma}\, e^{i\delta k_F x}
        \nonumber\\
    &&\qquad\qquad
        +\, R^{\dagger}_{1\sigma} L_{2\sigma}\, e^{-i(k_{F,1}+k_{F,2})x}
        + L^{\dagger}_{1\sigma} R_{2\sigma}\, e^{i(k_{F,1}+k_{F,2})x}
        + \text{H.c.}
    \Big].
\end{eqnarray}
Due to the mismatch between the Fermi momenta of the two legs, this single-particle tunneling rapidly oscillate. Consequently, these terms average to zero under coarse-graining and are effectively RG-irrelevant.

We further consider higher-order processes that emerge as the theory flows from the UV to the IR. Specifically, at second order in $t_{\perp}$, a two-particle pair-hopping process is generated:
\begin{eqnarray}
\mathcal{H}_{2p} &\sim& g_{2p} \int dx \left( c_{1\uparrow}^\dagger c_{1\downarrow}^\dagger c_{2\downarrow} c_{2\uparrow} + \text{H.c.} \right),
\end{eqnarray}
where the coupling constant scales as $g_{2p} \sim t_{\perp}^2$. Unlike single-particle tunneling, this pair-hopping operator remains commensurate. 
This operator can be expressed as:
\begin{eqnarray}
c_{1\uparrow}^\dagger c_{1\downarrow}^\dagger c_{2\downarrow} c_{2\uparrow} + \text{H.c.}
&\sim&
\cos(\sqrt{2}\theta_{-,c}-
\sqrt{2}\phi_{+,s}
).
\end{eqnarray}
Note that the spin-sector involves only the $\phi_{+,s}$, which is pinned as $0$ or $\pi$ in the iC-PDW phase. Thus the pair-hopping term is purely determined by the relative charge field $\theta_{-,c}$. 

The scaling dimension of the pair-hopping coupling $g_{2p}$ is given by:
\begin{eqnarray}
[g_{2p}] &=& 2 - \frac{1}{K_{-,c}}.
\end{eqnarray}
For $K_{-,c} > 1/2$, the pair-hopping term becomes relevant and additionally pins $\theta_{-,c}$ field. Consequently, the corresponding mode is gapped out, remaining a single gapless mode, resulting in a $c=1$ phase. Pinning of $\theta_{-,c}$ gaps out  charge 2e PDW correlations, as they involve the dual field, $\phi_{-,c}$ which highly fluctuates (see Eqs.~\ref{eq:pdw_inter_boson},~\ref{eq:pdw_intra_boson}). 

Instead, the dominant gapless mode is now a charge-$4e$ superconducting order, defined as $\Delta_{4e}\sim \mathcal{O}_{\mathrm{PDW},\perp}^2$, where
\begin{eqnarray}
\mathcal{O}_{\mathrm{PDW},\perp}
\sim
e^{i(k_{F,1}-k_{F,2})x}\,e^{-i\phi_{-,c}+i\theta_{+,c}}
+
e^{-i(k_{F,1}-k_{F,2})x}\,e^{i\phi_{-,c}+i\theta_{+,c}} .
\end{eqnarray}
Squaring this operator, one finds a uniform contribution that is independent of the relative charge field $\phi_{-,c}$,
\begin{eqnarray}
\Delta_{4e}
\sim
e^{2i\theta_{+,c}} .
\end{eqnarray}
As a result, while interlayer $2e$ PDW correlations are suppressed, the uniform charge-$4e$ correlations remain long-ranged decaying as 
\begin{eqnarray}
    \langle \Delta_{4e}^\dagger(0)
\Delta_{4e}(r)
\rangle 
\sim 
|r|^{-\frac{2}{K_{+,c}}}. 
\end{eqnarray}

\end{document}